\documentclass[prb,twocolumn,showpacs,floatfix,preprintnumbers,amsmath,amssymb,superscriptaddress]{revtex4}
\usepackage{graphicx}
\usepackage{dcolumn}
\usepackage{bm}
\usepackage{color}
\usepackage{color}
\usepackage[latin1]{inputenc}
\usepackage[urlcolor=blue]{hyperref}
\hypersetup{backref, colorlinks=true} 
\begin{document}

\preprint{APS/123-QED}

\title{The superconducting gaps in LiFeAs: Joint study of specific heat and ARPES}

\author{U.~Stockert}\email{ulrike.stockert@cpfs.mpg.de}\affiliation{Institute for Solid State Research, IFW Dresden, D-01171 Dresden, Germany}\affiliation{MPI for Chemical Physics of Solids, D-01187 Dresden, Germany}
\author{M.~Abdel-Hafiez}\affiliation{Institute for Solid State Research, IFW Dresden, D-01171 Dresden, Germany}
\author{D.~V.~Evtushinsky}\affiliation{Institute for Solid State Research, IFW Dresden, D-01171 Dresden, Germany}
\author{V. B. Zabolotnyy}\affiliation{Institute for Solid State Research, IFW Dresden, D-01171 Dresden, Germany}
\author{A.~U.~B.~Wolter}\affiliation{Institute for Solid State Research, IFW Dresden, D-01171 Dresden, Germany}
\author{S.~Wurmehl}\affiliation{Institute for Solid State Research, IFW Dresden, D-01171 Dresden, Germany}
\author{I.~Morozov}\affiliation{Moscow State University, Moscow 119991, Russia},\affiliation{Institute for Solid State Research, IFW Dresden, D-01171 Dresden, Germany}
\author{R.~Klingeler}\affiliation{Kirchhoff Institute for Physics, Heidelberg University, D-69120 Heidelberg}
\author{S. V. Borisenko}\affiliation{Institute for Solid State Research, IFW Dresden, D-01171 Dresden, Germany}
\author{B.~B\"uchner}\affiliation{Institute for Solid State Research, IFW Dresden, D-01171 Dresden, Germany}

\date{\today}

\begin{abstract}
We present specific heat, $c_P$, and ARPES data on single crystals
of the stoichiometric superconductor LiFeAs. A pronounced anomaly is
found in $c_P$ at the superconducting transition. The electronic
contribution can be described by two $s$-type energy gaps with
magnitudes of approximately $\Delta_1 = 1.2$~meV and $\Delta_2 =
2.6$~meV and a normal-state $\gamma$ coefficient of 10 mJ/mol K$^2$.
All these values are in remarkable agreement with ARPES results.
\end{abstract}

\maketitle

\subsection{Introduction}
The symmetry of the superconducting order parameter is one of the
basic characteristics of the superconducting state. In this respect,
the recently discovered pristine superconductor LiFeAs with a
$T_\mathrm{c}$ of approximately 18~K~\cite{F3-08-1, F3-08-2,
F3-08-3} plays a decisive role in elucidating the pairing mechanism
of the Cooper pairs and the nature of the superconducting state in
pnictide superconductors. In contrast to other pnictides,
superconductivity in LiFeAs evolves without additional doping, and
nesting between hole and electron pockets is very
poor.~\cite{F3-10-5} The evolution of spin-density wave (SDW) type
magnetic order, which is typically present in the vicinity of the
superconducting state in so-called '1111' and '122' pnictide
superconductors, is not observed in LiFeAs. Consequently, mediation
of superconductivity by antiferromagnetic spin fluctuations as
suggested for other pnictides~\cite{Theorie-1, Theorie-2} is
unlikely. Remarkably, there is evidence both from theory and
experiments for almost ferromagnetic fluctuations which drive an
instability toward spin-triplet p-wave
superconductivity.~\cite{Brydon2010, Baek2010}

Experimental investigations on the structure and magnitude of the
superconducting gaps in LiFeAs by means of bulk specific heat data
as well as by angle resolved photoemission spectroscopy (ARPES) are
of great interest. Previous specific heat data obtained on an
assembly of tiny LiFeAs single crystals are in agreement with the
presence of two isotropic gaps of 0.7 meV and 2.5 meV, although the
presence of nodes could not be ruled out.~\cite{F3-10-1} Similarly,
magnetization measurements on polycrystals suggest two $s$-type gaps
of 0.6 meV and 3.3 meV, but do not exclude gap nodes.~\cite{F3-10-4}
Recent magnetization data of single crystals revealed two gaps of
approximately 1.3~meV and 2.9 meV.~\cite{Song2010} Measurements of
the London penetration depth of single crystals are in line with
nodeless superconductivity and two gaps of 1.7~meV and
2.9~meV.~\cite{Kim2010, Imai2010} First ARPES measurements suggest
the presence of two gaps of 1.5 meV and 2.5 meV as
well~\cite{F3-10-5}, with the larger one being in reasonable
agreement with the results of the data analysis presented
in~\onlinecite{F3-10-2}.

In this manuscript we study the superconducting energy gaps of
LiFeAs single crystals by two complementary experimental methods,
specific heat measurements and ARPES. The specific heat is sensitive
to the bulk and gives direct access to the entropy of Cooper pair
breaking, which is determined by the superconducting gap structure.
In turn, ARPES allows for probing the momentum-resolved
superconducting gap. From our results we can exclude the possibility
of $d$-wave pairing in LiFeAs. Instead, both methods are in line
with the existence of at least two $s$-type energy gaps for
different Fermi surface sheets of LiFeAs with magnitudes of
approximately $\Delta_1 = 1.2$~meV and $\Delta_2 = 2.6$~meV.


\subsection{Experimental methods}

Single crystals of LiFeAs have been prepared by a self-flux method
described in detail in Ref.~\onlinecite{F3-10-6}. Specific heat data
were obtained by a relaxation technique in a PPMS (Quantum Design)
on a sample with a mass of 2.4~mg. Measurements of the magnetic
susceptibility in a field of 2 mT (MPMS-XL from Quantum Design)
confirmed that this sample has a magnetic $T_\mathrm{c}^\chi$ of
16.9~K, similar to other crystals from the same
batch.~\cite{F3-10-6} After correction for demagnetization effects
using an ellipsoid approximation for the sample shape, the
superconducting volume fraction was estimated to 0.91. The
difference to 100~\% is within the error of demagnetisation and may
be due to deviations of the sample from an ellipsoid and/or due to a
small non-superconducting phase. Photoemission experiments have been
carried out using the synchrotron radiation from the BESSY storage
ring. The end-station ``1-cubed ARPES'' is equipped with a $^3$He
cryostat which allows to collect angle-resolved spectra at
temperatures below 1~K. All single crystals have been cleaved in UHV
exposing the mirror-like surfaces.


\subsection{Results}

\begin{figure}[tb]
\begin{center}
\includegraphics[width=0.95\columnwidth]{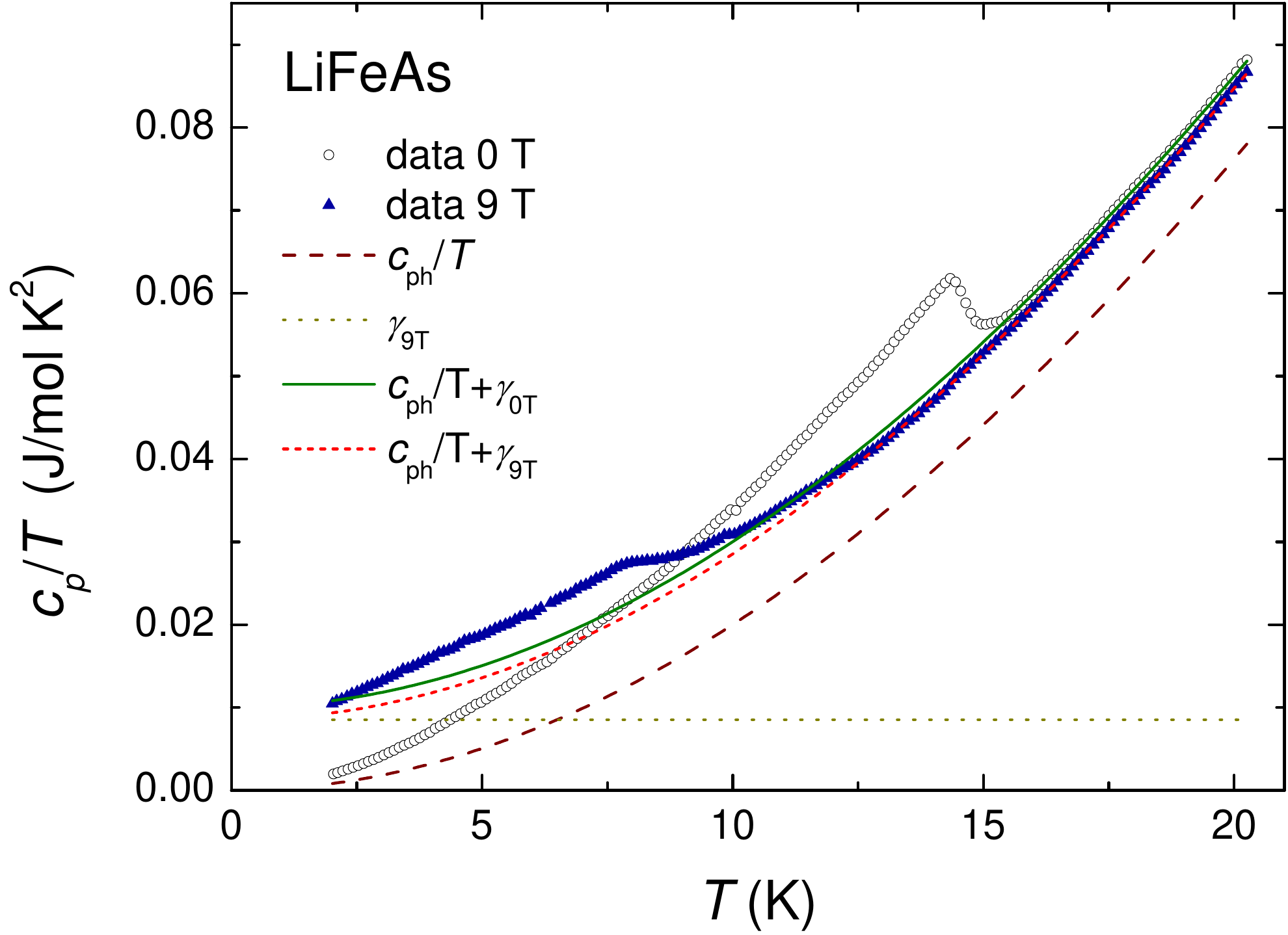}
\end{center}
\caption{The temperature dependence of the specific heat of LiFeAs
is shown as $c_p/T$ vs.~$T$ in zero magnetic field and in a field $B
\parallel c$ of 9 T. The phonon and electron contributions
$c_\mathrm{ph}$ and $c_\textup{el-n\, 9T}$ have been determined from
a fit of the 9~T data as explained in the text. In addition, the
fitted total $c_p/T$ for 0~T and 9~T are shown. \label{cp}}
\end{figure}

The temperature dependence of the specific heat $c_p$ of a LiFeAs
single crystal is shown in Fig.~\ref{cp} as $c_p/T$ vs.~$T$. In zero
magnetic field a clear anomaly is observed around 15 K which is
attributed to the superconducting phase transition. By applying a
magnetic field $B = \mu_0H \parallel c$ of 9~T the anomaly is
shifted to lower temperatures and reduced in height.

In order to determine the specific heat related to the
superconducting phase transition we need to estimate the phonon
($c_\mathrm{ph}$) and electron ($c_\textup{el-n}$) contributions to
$c_p$ in the normal state. At low $T$, $c_\textup{el-n}$ behaves
linear in $T$, while $c_\mathrm{ph}$ varies as $c_\mathrm{ph}\propto
T^3$. However, for LiFeAs the onset of superconductivity limits the
fitting range towards low $T$. In order to improve the reliability
at higher $T$, we used a second term of the harmonic-lattice
approximation, i.e. $c_\textup{el-n}+c_\mathrm{ph} = \gamma T +
\beta_3 T^3 + \beta_5 T^5$. The results of a fit of the 9\,T data
between 13\,K and 20\,K are shown as lines in Fig.~\ref{cp}. From
the fitting parameters we calculated a Debye temperature of 310~K
and a Sommerfeld coefficient of $\gamma _\mathrm{9T} =
8.5$~mJ/mol~K$^2$. Below 13\,K the 9\,T data deviate slightly from
the fit, although the superconducting transition is observed only
around 8\,K. This may be due to a tiny part of the sample with
different orientation $B
\parallel ab$. For this field direction the superconducting
transition in 9\,T is known to take place at
13\,K.~\cite{kurita2010} The data above 13\,K are well described by
the fit.

The zero field specific heat is shifted by a constant with respect
to the 9\,T curve above $T_\mathrm{c}$, which is attributed to a
field-dependent electronic specific heat. From our data we obtain a
zero-field Sommerfeld coefficient \mbox{$\gamma_\mathrm{0T} =
\gamma_\mathrm{n} = 10.0$\,mJ/mol K}. This value is about half the
one determined in previous specific heat studies on polycrystalline
LiFeAs~\cite{F3-09-1} and on an assembly of small single
crystals.~\cite{F3-10-1} The differing $\gamma$ coefficient may
arise from differences in the sample quality. Resistivity
measurements on a crystal from the same batch as the one
investigated here revealed a very low residual resistivity of only
15.2~$\mu \Omega$cm and a large residual resistivity ratio RRR =
$\rho_{300\,\mathrm{K}}/\rho_{0\,\mathrm{K}}$ of
38.~\cite{Heyer2010} This confirms the high quality of our samples.
By contrast, the polycrystalline LiFeAs investigated in
Ref.~\onlinecite{F3-09-1} exhibits a much larger residual
resistivity of approximately 2.5~m$\Omega$cm with RRR $\approx 10$.
This may be an indication for the presence of impurities, which can
give rise to additional contributions to the specific heat.

\begin{figure}[tb]
\begin{center}
\includegraphics[width=0.95\columnwidth]{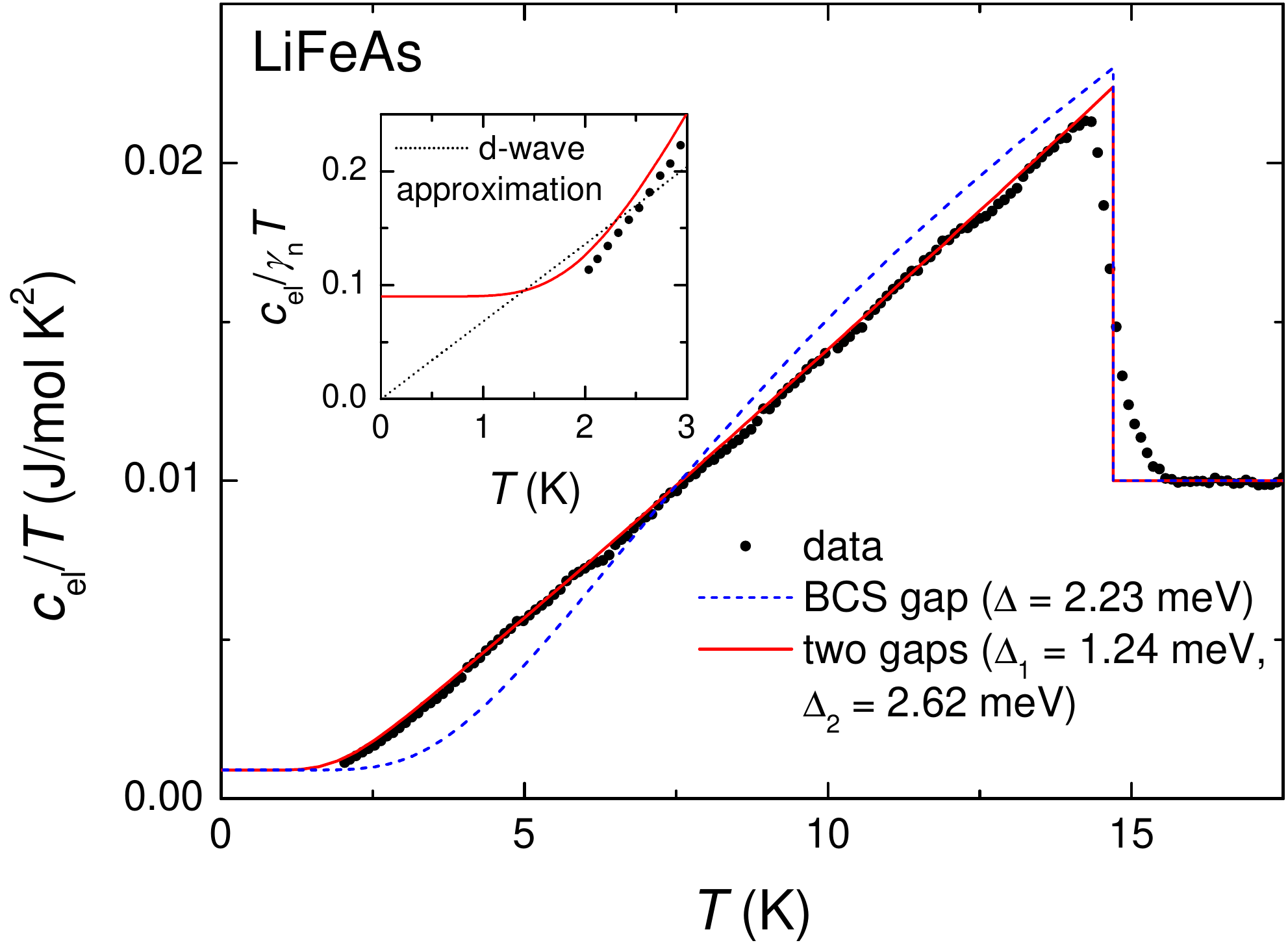}
\end{center}
\caption{The main plot shows the electronic contribution to the
specific heat of LiFeAs in zero field as $c_\mathrm{el}/T$ vs.~$T$.
The dotted line corresponds to a single-gap BCS curve taking into
account a residual nonsuperconducting contribution with
$\gamma_\mathrm{res}/\gamma_\mathrm{n} = 0.09$. The red curve is a
two-gap model of the data assuming likewise
$\gamma_\mathrm{res}/\gamma_\mathrm{n} = 0.09$. The inset shows the
low-$T$ data as $c_\mathrm{el}/\gamma_\mathrm{n}T$ vs.~$T$ on a
larger scale. In addition, an estimate for the low-$T$
$c_\mathrm{el}$ expected in case of a d-wave order parameter is
plotted. \label{gap}}
\end{figure}

Fig.\,\ref{gap} shows the temperature dependence of the electronic
contribution to the specific heat $c_\mathrm{el}$ in zero field
determined by subtracting $c_\mathrm{ph}$. From an
entropy-con\-serving construction we find a superconducting
transition temperature of $T_\mathrm{c} = 14.7$\,K, which is
somewhat smaller than the magnetic $T_\mathrm{c}^\chi$ of the
sample. This difference between the thermodynamic and magnetic
$T_\mathrm{c}$ has been also found in previous studies of
LiFeAs.~\cite{F3-10-1, F3-09-1} Of particular interest in this
context is a recent investigation of single-crystalline
LiFeAs:~\cite{kurita2010} The onset value of $T_\mathrm{c}$
determined by dc susceptibility in 1\,mT was found to be about 17~K.
However, the extrapolation of the field dependence $T_\mathrm{c}(B)$
determined by magnetic torque measurements on a small piece from the
same crystal yielded a zero-field $T_\mathrm{c}$ of 15.5~K.
Likewise, ac-susceptibility data for $B\gtrsim 0.5$~T on crystals
from the same batch extrapolate to a lower $T_\mathrm{c}$ than the
values measured in very small fields. Therefore, the difference
between the low-field $T_\mathrm{c}^\chi$ and the bulk
$T_\mathrm{c}$ determined, e.g., by specific heat appears inherent
to LiFeAs.

The jump height of $c_\mathrm{el}/T$ at $T_\mathrm{c}$ amounts to
1.24 $\gamma _n$, which is slightly lower than the BCS value of 1.43
$\gamma _n$. In addition, the almost linear temperature dependence
of $c_\mathrm{el}/T$ indicates, that the specific heat data cannot
be described by a single BCS gap. In order to illustrate this we
show a theoretical BCS curve with $\Delta = 1.764\,
k_{\mathrm{B}}T_\mathrm{c} = 2.23$~meV in Fig.~\ref{gap}, where we
account for a small fraction $\gamma_\mathrm{res}/\gamma_n = 0.09$
of normal electrons as justified below. Systematic deviations of the
single-gap fit from the data are observed in the whole temperature
range below $T_\mathrm{c}$. Since a single gap cannot describe the
data, we applied a phenomenological two-gap model developed for the
specific heat of MgB$_2$.~\cite{SC-01-1} For this purpose we
calculated theoretical curves $c_\mathrm{el}/T$ vs.~$T$ for a large
range of the free parameters $\Delta_1$, $\Delta_2$,
$\gamma_\mathrm{res}$, and the weight of the gaps
$w_{\Delta2/\Delta1}$. We then calculated the differences $d_i$
between these curves and the data point by point and used the sum of
the squares $\Sigma d_i^2$ as criteria for the model quality. The
best description of the data is obtained for $\Delta_1 = 1.44$~meV,
$\Delta_2 = 2.74$~meV, $\gamma_\mathrm{res}/\gamma_\mathrm{n} =
0.13$, and $w_{\Delta2/\Delta1} = 1.18$. This result, however, is
not realistic: The measured data exclude values larger than 0.11 for
$\gamma_\mathrm{res}/\gamma_\mathrm{n}$. This is seen from the inset
of Fig.~\ref{gap}, which shows the low-temperature ratio
$c_{\mathrm{el}}/\gamma_\mathrm{n} T$. It reaches a value of 0.11
around 2~K. In view of the systematic decrease of the data with
decreasing temperature it is likely, that the ratio
$\gamma_\mathrm{res}/\gamma_\mathrm{n}$ is even lower. Therefore, we
take the normal state contribution determined from the magnetic
susceptibility of the same sample to estimate
$\gamma_\mathrm{res}/\gamma_\mathrm{n} = 0.09$. By doing so, the
best description of the data is obtained for $\Delta_1 = 1.24$~meV,
$\Delta_2 = 2.62$~meV, and $w_{\Delta2/\Delta1} = 1.53$. The
corresponding calculated specific heat is shown as a red line in
Fig.~\ref{gap}. It is in a very good agreement with the data in the
whole temperature range. Still, a close look to the low-$T$ part
(cf.~inset of Fig.~\ref{gap}) reveals systematic deviations from the
data, which suggests an even lower $\gamma_\mathrm{res}$.

Although the data are best reproduced by the parameters given above,
they can be described with similar accuracy for a considerable range
of gaps. For an estimate we consider all curves for which the
deviation from the data $\Sigma d_i^2$ is at most 2 times the value
for the best curve with $\gamma_\mathrm{res}/\gamma_\mathrm{n} =
0.09$. As an additional constraint we assume
\mbox{$\gamma_\mathrm{res}/\gamma_\mathrm{n} \leq 0.11$} in
agreement with the data. Thus, we obtain \mbox{$\Delta_1 =
(0.93-1.67)$~meV, $\Delta_2 = (2.40-3.24)$~meV},
\mbox{$\gamma_\mathrm{res}/\gamma_\mathrm{n} = 0.04-0.11$}, and
\mbox{$w_{\Delta2/\Delta1} = 0.46-3.45$}.

\begin{figure}[tb]
\begin{center}
\includegraphics[width=0.98\columnwidth]{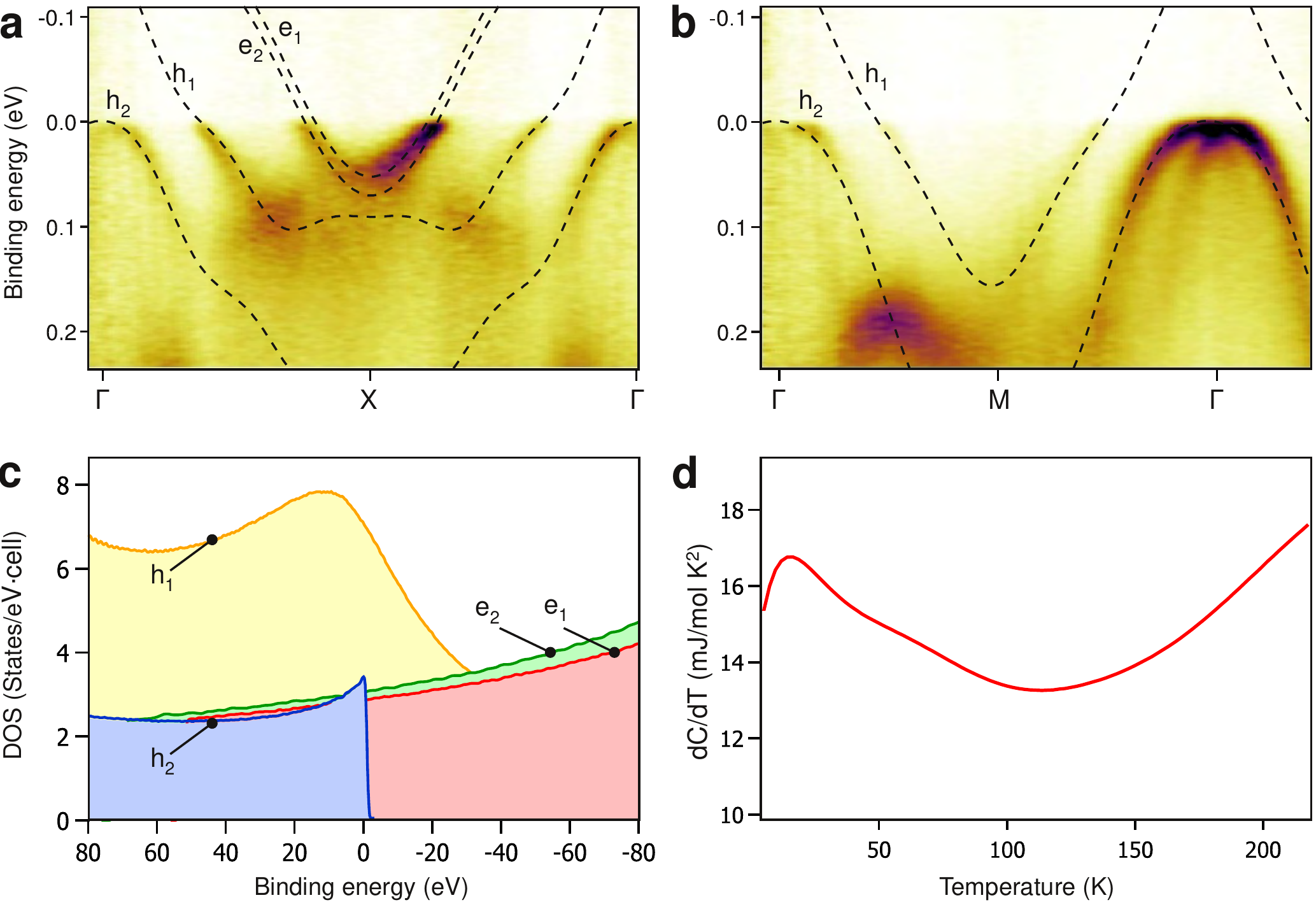}
\end{center}
\caption{ (a), (b) Photoemission intensity in the high symmetry cuts
together with the fitted tight-binding dispersions. e$_1$, e$_2$ are
the two bands forming the electron pockets at the X point of the BZ.
h$_1$, h$_2$ are the outer and the inner hole bands centered at the
$\Gamma$ point. (c) Band contribution to the DOS derived from the
fitted quasiparticle dispersions. (d) Temperature dependent
Sommerfeld coefficient determined as $\gamma(T) =
\mathrm{d}c_\mathrm{el}/\mathrm{d}T$. \label{Arpes1}}
\end{figure}

The overall shape of the superconducting anomaly presented here is
similar to the one obtained recently on a cluster of tiny single
crystals of LiFeAs.~\cite{F3-10-1} However, the magnitudes of the
gaps are somewhat larger, in particular for the smaller gap of
1.24\,meV compared to 0.7~meV obtained in Ref.~\onlinecite{F3-10-1}.
This may be due to the limited resolution of the data in
Ref.~\onlinecite{F3-10-1}, which leaves a considerable uncertainty
for $\gamma _\mathrm{res}$. In addition, the magnitude of the
smaller gap $\Delta_1$ itself is very sensitive to the specific heat
at low $T$. The high resolution of our data down to 2~K allows for a
very reliable estimate of both $\gamma _\mathrm{res}$ and
$\Delta_1$. In Ref.~\onlinecite{F3-10-1} even a $d$-wave order
parameter could not be ruled out. In this case one should find a
linear-in-$T$ behavior of $c_\mathrm{el}/T$ for $T \ll
T_\mathrm{c}$. Since our measurement range is limited to $T >
0.14$\,$T_\mathrm{c}$, we cannot directly exclude the presence of
line-nodes of the gap function, either. However, the quality of our
data and the low value at 2~K render the presence of line nodes of
the gap function very unlikely. This is demonstrated by the expected
low-$T$ behavior for a $d$-wave order parameter estimated as
described in Ref.~\onlinecite{F3-10-1} and assuming
$\gamma_\mathrm{res} = 0$. It cannot be brought into agreement with
the measured data. Taking into account in addition recent ARPES
results suggesting a nearly isotropic gap for each Fermi surface
sheet, \cite{F3-10-5} a $d$-wave order parameter is excluded from
consideration. \vspace{0.3cm}

Owing to its ability to resolve both momentum and energy of the
electronic states, ARPES can provide a complete picture of the
electronic band dispersion and the momentum-dependent
superconducting gap. Therefore, it is interesting to compare the
thermodynamic properties measured directly with those calculated
from photoemission data. In particular, one can quite easily extract
the value of the Sommerfeld coefficient $\gamma_\mathrm{n}$. This
parameter determines the heat capacity in the normal state, and,
along with the superconducting gap values and the assumption of BCS
pairing, the thermodynamic properties in the superconducting state.
For this purpose the photoemission intensity of the LiFeAs has been
mapped in more than one Brillouin zone (BZ) and fitted with the
standard tight-binding formula.~\cite{Kordyuk2003, Inosov2008} To
demonstrate the agreement between the ARPES raw data and the
obtained tight-binding fits, in Fig.~\ref{Arpes1}a and b we show two
high symmetry cuts along $\Gamma$-X-$\Gamma$ and $\Gamma$-M-$\Gamma$
directions with the fits to the renormalized quasiparticle
dispersions superimposed over the ARPES data. As one may see from
the derived density of states (DOS) (Fig.~\ref{Arpes1}c), the major
contribution to the heat capacity must be due to the outer hole
band. Another interesting observation is a quite pronounced
variation in the DOS at the Fermi level, which may result in
deviations from the linear temperature dependence of the electronic
heat capacity. To check to which extent this applies to the current
case, instead of using a standard text-book expression for the
electronic heat capacity $c_\mathrm{el} \propto
\mathcal{D}(E_{\mathrm{F}})k_{\mathrm{B}}T$, we have made an
estimate based on a more general expression, which for a single
quasiparticle band with dispersion $\mathcal{E}_{k_{i}}$ reads as
$c_M = 2 N_{\mathrm{A}} \frac{\partial}{\partial T}\langle f
(\mathcal{E}_{k_{i}},T)\mathcal{E}_{k_{i}}\rangle _\textup{BZ}$.
Here $\langle...\rangle _\textup{BZ}$ denotes an average over the
Brillouin zone and $N_{\mathrm{A}}$ is the Avogadro constant.
Indeed, the value of $\mathrm{d}c_\mathrm{el}/\mathrm{d}T$, that in
case of strict temperature linearity defines the Sommerfeld
coefficient $\gamma_{n}$, varies from 13 to about 17\,mJ/mol
K$^{2}$, which is in a relatively good agreement with the direct
measurements resulting in $\gamma _{n} \approx 10$ mJ/mol\,K$^2$.
This variation in the $\mathrm{d}c_\mathrm{el}/\mathrm{d}T$ is also
likely to account for some variation in the $\gamma_{n}$ extracted
from the thermodynamic measurements by different authors.

It is remarkable that the thermodynamic gap values determined in the
present study are in excellent agreement with the ARPES leading-edge
gaps reported in Ref.~\onlinecite{F3-10-5} (1.5 meV and 2.5 meV).
However, the absolute values of the actual gaps, which can be
derived from the ARPES data after a more rigorous analysis, are
usually slightly higher than the leading-edge ones. The resulting
discrepancy (of the order of 0.5 meV) between the absolute gap
values derived from ARPES and in the present study could probably be
explained by the difference in thermodynamic and magnetic
$T_{\mathrm{c}}$ mentioned before. In any case, the ratio of the
ARPES leading-edge gaps reproduces the ratio of the actual gap
values quite accurately, and this is in close correspondence with
the ones discussed here. Moreover, a more detailed investigation of
the superconducting gaps in LiFeAs~\cite{Borisenko-prep} indicates
that the value of the gap supported by the band h2 is indeed
comparable to that of the large hole Fermi surface, made by the band
h1 (see Fig. 3a). In the light of the results presented in Fig.3c,
it is important to establish the fact that the smaller gap
corresponds to the hole-like Fermi surfaces centered around the
Gamma point, while the larger one corresponds to the electron-like
Fermi surfaces localized around the corners of the Brillouin zone.
This knowledge can help to identify the symmetry of the order
parameter in iron pnictides in more details.




\subsection{Summary}
In summary, both thermodynamic and spectroscopic experiments on
LiFeAs render a nodal gap very unlikely, and equivocally speak in
favor of a strong variation of the gap magnitude between different
electronic bands, from about 1.2~meV to 2.6~meV. The general
agreement of such complementary probes within band picture
emphasizes the robustness of the conclusions drawn. The multigap
behavior of LiFeAs established above, is in line with two gaps found
in many other iron arsenides.

\begin{acknowledgments}
Work was supported by the Deutsche Forschungsgemeinschaft through
the Priority Program SPP1458 (BE1749/13). I.M. acknowledges support
from the Ministry of Science and Education of the Russian Federation
under State contract P-279 and by RFBR-DFG (Project No.
10-03-91334).

\end{acknowledgments}


\bibliographystyle{PRBstyle}
\bibliography{LiFeAs_gap}

\end{document}